\documentclass[a4paper,12pt]{article}
\pdfoutput=1
\usepackage{abstract} 
\usepackage{amsmath}
\RequirePackage{graphicx}
\RequirePackage{geometry}   
\RequirePackage{fancyhdr} 
\RequirePackage{enumitem}  
\usepackage{caption,subcaption}
\usepackage{tikz}
\usepackage{multirow}  
\usepackage{tabularx} 
\usepackage{float}
\usepackage{makecell} 
\usepackage{multicol}
\usepackage{color}
\usepackage{physics}
\usepackage[style=gb7714-2015,backend=bibtex]{biblatex}
\usepackage{hyperref}
\hypersetup{
	colorlinks=true,
	linkcolor=blue,
	urlcolor=blue,
	filecolor=blue,
	citecolor=cyan,
}
\usepackage{indentfirst}
\usetikzlibrary{arrows,shapes,chains}
 \selectfont
\begin{document}

	
	\renewcommand\thesubsection{}  
	
	\renewcommand{\abstractname}{} 
	
	\title{The Rotation Curve and Spiral Structure of Milky Way from the Hydrogen 21-cm Line Detection with Campus Radio Telescope \vspace{-2em}}
	
	\date{}
	\maketitle
	\thispagestyle{empty}
	\vspace{-2em}
	\centerline{\large {Hong-Yu Li$^1$}\quad\quad \large{Jia-Hao Hu$^2$}}

	\centerline{Department of Astronomy, School of Physics, Huazhong University of Science and Technology,}
    \centerline{Wuhan, 430074, People's Republic of China}
	\vspace{1em}
	\centerline{\textbf{\large{ABSTRACT}}}

	 The rotation curve is significant to research on dark matter and the structure of the Milky Way. But it is a great challenge to measure rotation curve accurately, and different ways obtain distinct results. In this work, we use a DIY small radio telescope to carry out hydrogen 21-cm line observations on campus, and calculate the rotation curve of inner disk by tangent method based on the model of Przemek et al. (the rotation speed of the Sun $v(R_0)=233.6\pm2.8$ km/s, and distance to the Galactic center $R_0=8.122\pm0.031$ kpc)\autocite{mroz2019rotation}. Furthermore, we obtain the distribution of spiral arms in the Milky Way with our data by adopting different rotation curve models. We also compare our rotation curve result with previous works, analyse the possible measurement error in the tangent method, and discuss the consistence of our result with others. \par 
	 \par
	\begin{multicols}{2}
		\section{Introduction}
		The rotation curve is an important tool to describe the dynamic properties of Milky Way Galaxy. Variety of methods are used to develop the rotation curve, the key is to observe the positions and velocities of numerous objects on galactic plane. For fixed stars, the Doppler redshift of the characteristic spectral lines produce radial velocity and the proper motions produce transverse velocity.  Przemek et al. observed Classical Cepheids and obtained their distance to Sun based on period-luminosity relations \autocite{mroz2019rotation},Reid et al. observed the parallax of stars from High-mass Star Forming Regions\autocite{reid2014trigonometric},Eilers et al. observed a large number of luminous red giant stars' parallax and calculated their position by trigonometric parallax method.\autocite{eilers2019circular}Clemens et al. used the same tangent method as our work, observed the CO emission line and measured the rotation curve of the inner disk. \autocite{clemens1985massachusetts}Nowadays, more and more universities set up department of astronomy, but astronomy education, especially the astronomical observation and experiment is in deficiency. In this study, we utilized the small radio telescope at our school to observe the 21cm neutral hydrogen spectral line in the Milky Way galaxy. By employing the tangent method, we obtained the rotation curve of the inner disk and also identified the local spiral arm structure of the Milky Way. We provided a comprehensive solution for radio astronomy experiments at universities. The relevant results are in good agreement with those in the literature, demonstrating the feasibility of this approach for astronomical exploration and talent cultivation. Furthermore, the 1.4GHz frequency band, where the 21cm neutral hydrogen spectral line resides, is an internationally protected frequency band, which also provides conditions for astronomical observations within the university. \par 
		Our observations tracked the movement of material around the center of the Milky Way by detecting the 1420MHz hyperfine structure spectral line of the widespread HI atoms in the Galactic plane. By analyzing the emission spectra of HI atoms, we obtained information on red shift and the relative motion between the material and the observer. This was complemented by a series of corrections for relative motions to derive the linear velocities of material rotating around the Galactic center at different radii. Combining the rotation curve model with geometric relationships, we further analyzed the emission spectra of HI atoms to derive the regions of high-density neutral hydrogen distribution in the Milky Way galaxy. From these regions, the spiral arm structure of the Milky Way galaxy was clearly discernible. We compared our observed rotation curve model with previous models, analyzed potential errors in neutral hydrogen tangent method measurements, and provided a brief theoretical justification.Furthermore, we calculated the distribution of Milky Way spiral arms using different rotation curve models and compared them with each other. We evaluated the degree of support for various rotation curve models based on the data processing procedure and the output results of the images.\par 
		In Section 2 we will briefly introduce the undergraduate innovation and research background, the DIY small radio telescope, and examples of observed spectra. In Section 3 we will focus on introducing the basic theory, main tools, and methods of our data processing, along with presenting the calculated rotation curve and spiral arm distribution map. These results demonstrate the feasibility of on-campus astronomical observation experiments. In Section 4 we emphasizes comparing our results with those of previous researchers, analyzing potential systematic errors in the observation method, discussing issues and phenomena in the data processing process, and attempting to provide explanations in terms of physical imagery. In Section 5 we will briefly summarize the conclusions obtained from the observations and subsequent processing.\par 
		\section{Observation and Data Reduction}
		\subsection{\normalsize{Background}}
		The hydrogen 21-cm line arises from the transition between two hyperfine energy levels of the ground state hydrogen atom. It falls within the microwave frequency range, easily observable as it can penetrate the atmosphere, marking the first spectral line observed in radio astronomy. Since its first observation in 1951, advancements in electronic technology have made the observation and study of hydrogen 21-cm line achievable with small radio telescopes, not necessarily need the large ones. In this work, supported by a university student innovation and entrepreneurship training program, we constructed a small radio telescope and conducted hydrogen spectral line data collection on campus using SDRSharp software, subsequently analyzing the 21cm hydrogen spectral line data.\par 
		\subsection{\normalsize{Composition of our Radio Telescope}}
\begin{figure}[H]
		\begin{minipage}[b]{0.45\textwidth}
			\centering
			\includegraphics[width=\textwidth]{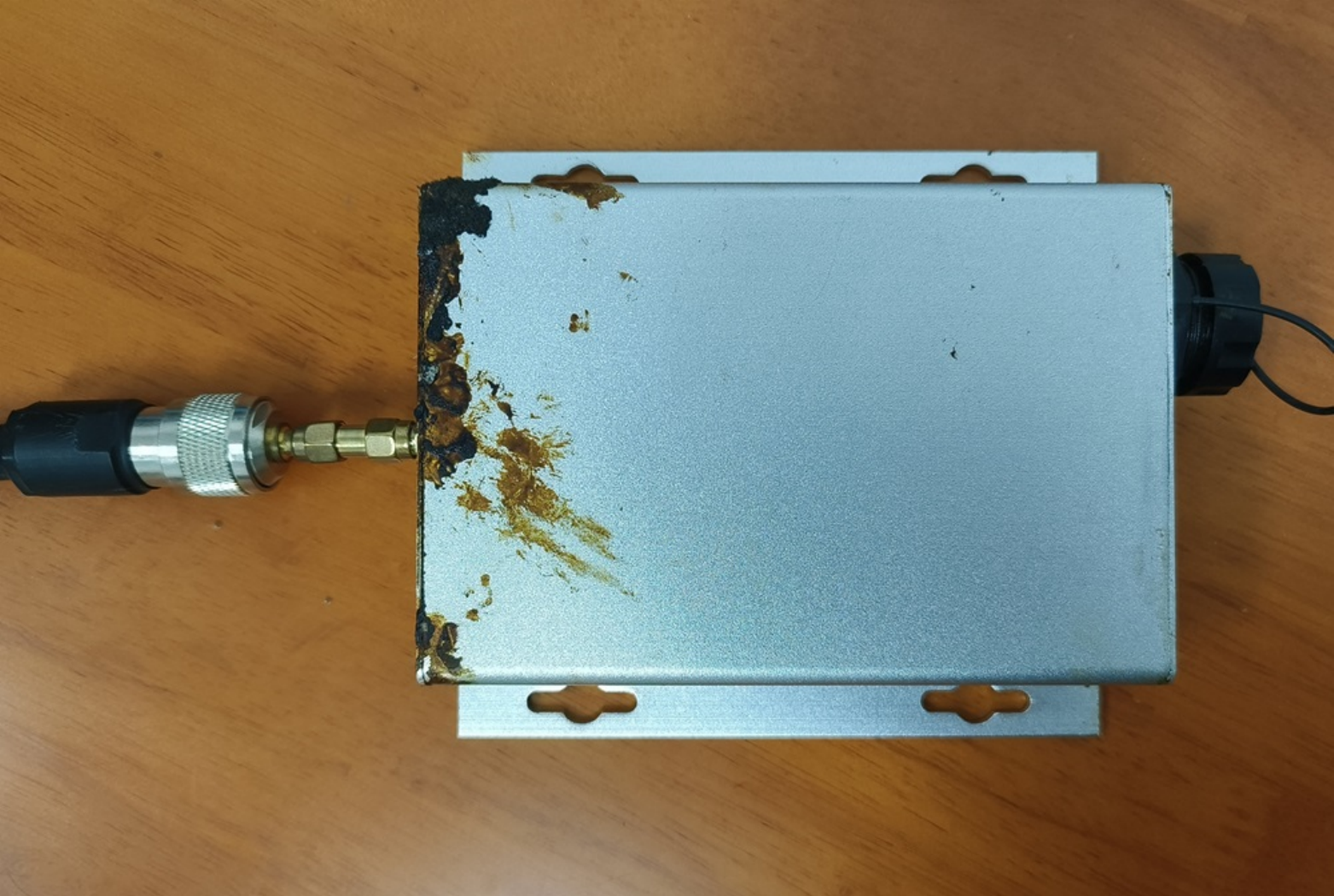}
			\caption{receiver integrated into an existing metal box, connect antenna on the left side and computer on the right side}
			\label{fig:jc}
		\end{minipage}
		\begin{minipage}[b]{0.45\textwidth}
			\centering
			\includegraphics[width=0.8\textwidth]{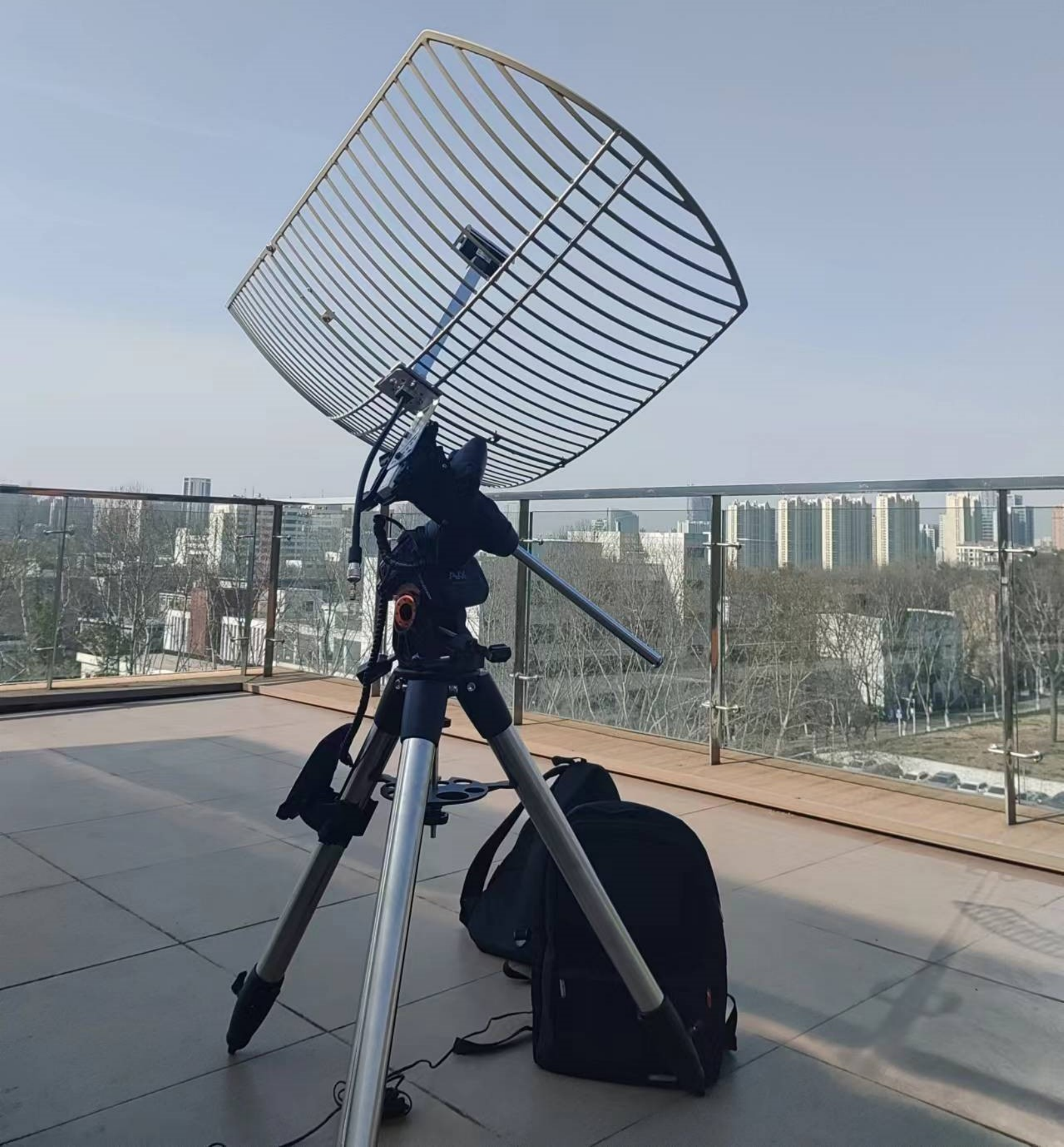}
			\caption{telescope installation}
			\label{fig:zuhe}
		\end{minipage}
	\end{figure}

		The fundamental composition of our radio telescope is shown in Figure \ref{fig:syt} ,the physical image is shown in Figure \ref{fig:swt}.To reduce electromagnetic interference, a receiver integrated into an existing metal box is used during actual observations, as Figure \ref{fig:jc} show.Simultaneously, the radio telescope needs to be mounted on an equatorial mount with Goto functionality in order to be used as a pointing system, as Figure \ref{fig:zuhe} show.\\\\\\
	\end{multicols}
	
	\tikzstyle{block} = [rectangle, draw, text width=6em, text centered, rounded corners, minimum height=3em]
	\tikzstyle{line} = [draw, -latex']
	
	\begin{figure}[H]
		\centering
		\begin{tikzpicture}[node distance = 3cm, auto]
			\node [block] (antenna) {grid antenna};
			\node [block, right of=antenna] (lna) {low noise amplifier(LNA)};
			\node [block, right of=lna] (filter) {wave filter};
			\node [block, right of=filter] (receiver) {receiver};
			\node [block, right of=receiver] (computer) {computer};
			\path [line] (antenna) -- (lna);
			\path [line] (lna) -- (filter);
			\path [line] (filter) -- (receiver);
			\path [line] (receiver) -- (computer);
		\end{tikzpicture}
		\caption{Composition of Radio Telescope}
		\label{fig:syt}
	\end{figure}
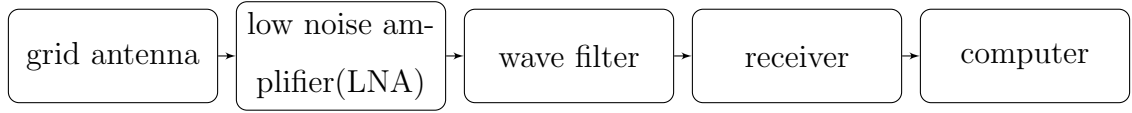
	\begin{figure}[H]
		\centering
		\begin{minipage}[b]{0.3\textwidth}
			\centering
			\includegraphics[width=\textwidth]{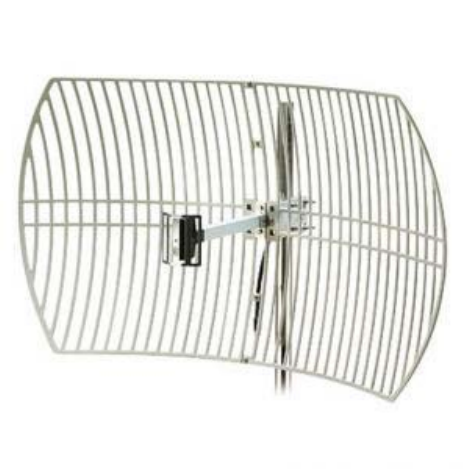}
			\caption{grid antenna}
		\end{minipage}
		\begin{minipage}[b]{0.65\textwidth}
			\centering
			\includegraphics[width=0.6\textwidth]{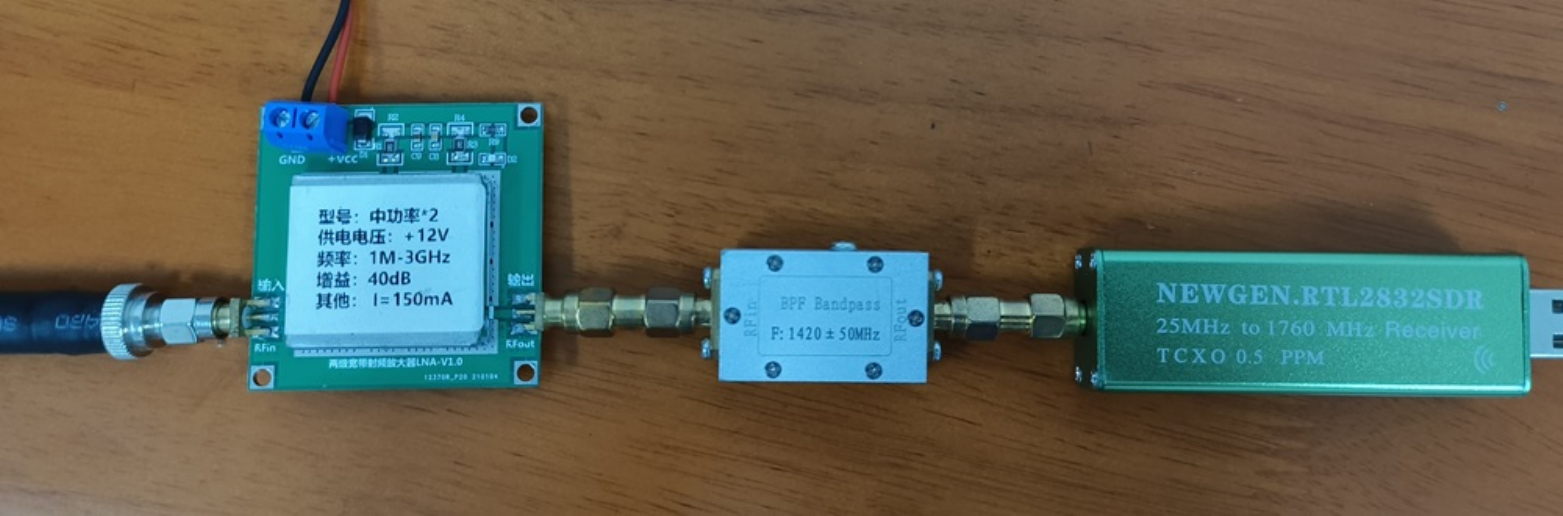}
			\caption{the physical image, from left to right: antenna cable, LNA, wave filter, receiver, and connect computer on the right side}
			\label{fig:swt}
		\end{minipage}
	\end{figure}

	\begin{multicols}{2}
		\subsection{\normalsize{Data Extraction}}
\begin{figure}[H]
		\centering
		\begin{minipage}[b]{0.45\textwidth}
			\centering
			\includegraphics[width=\textwidth]{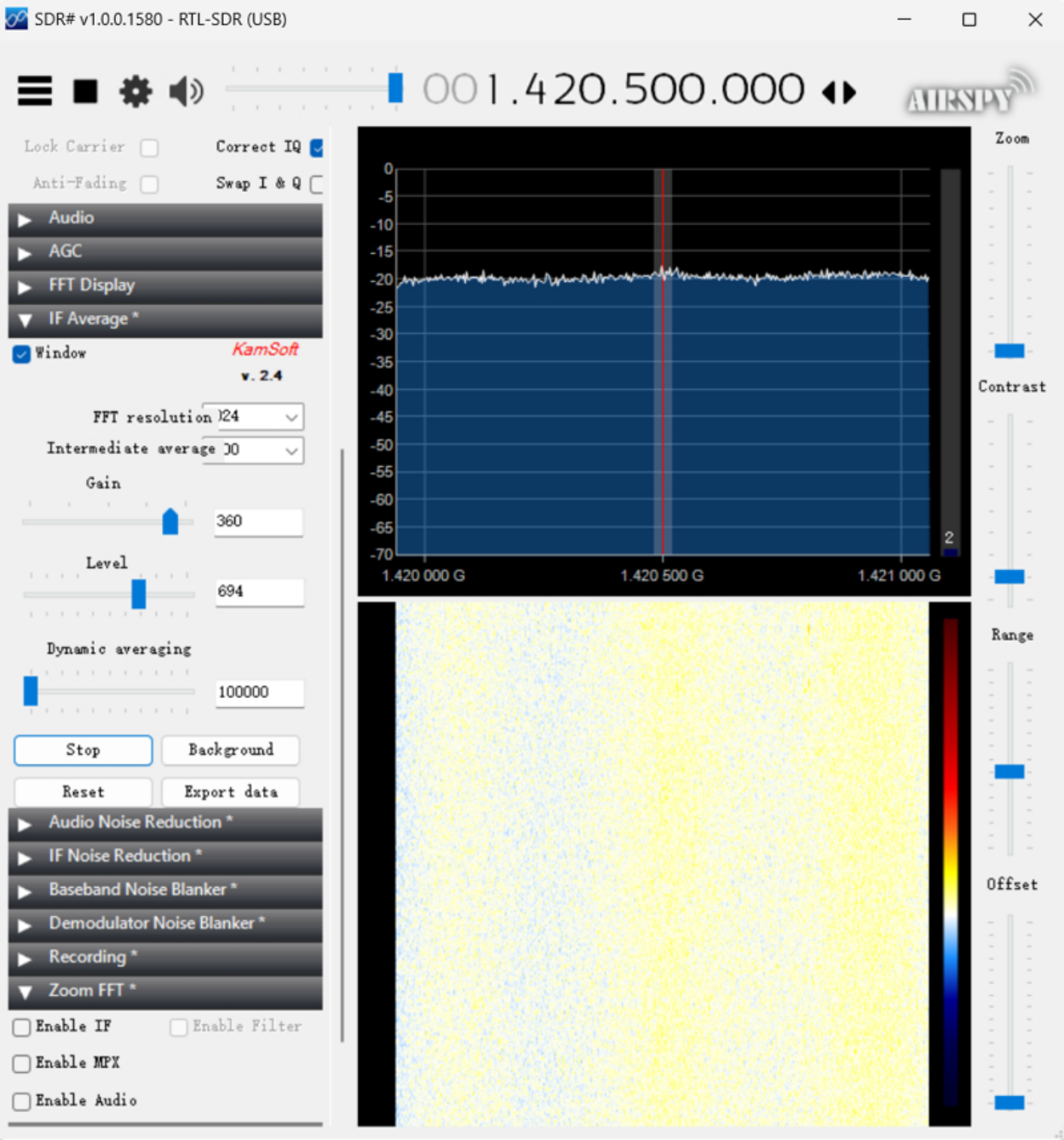}
			\caption{SDRSharp main window}
			\label{fig:sdr2}
		\end{minipage}
		\begin{minipage}[b]{0.45\textwidth}
			\centering
			\includegraphics[width=\textwidth]{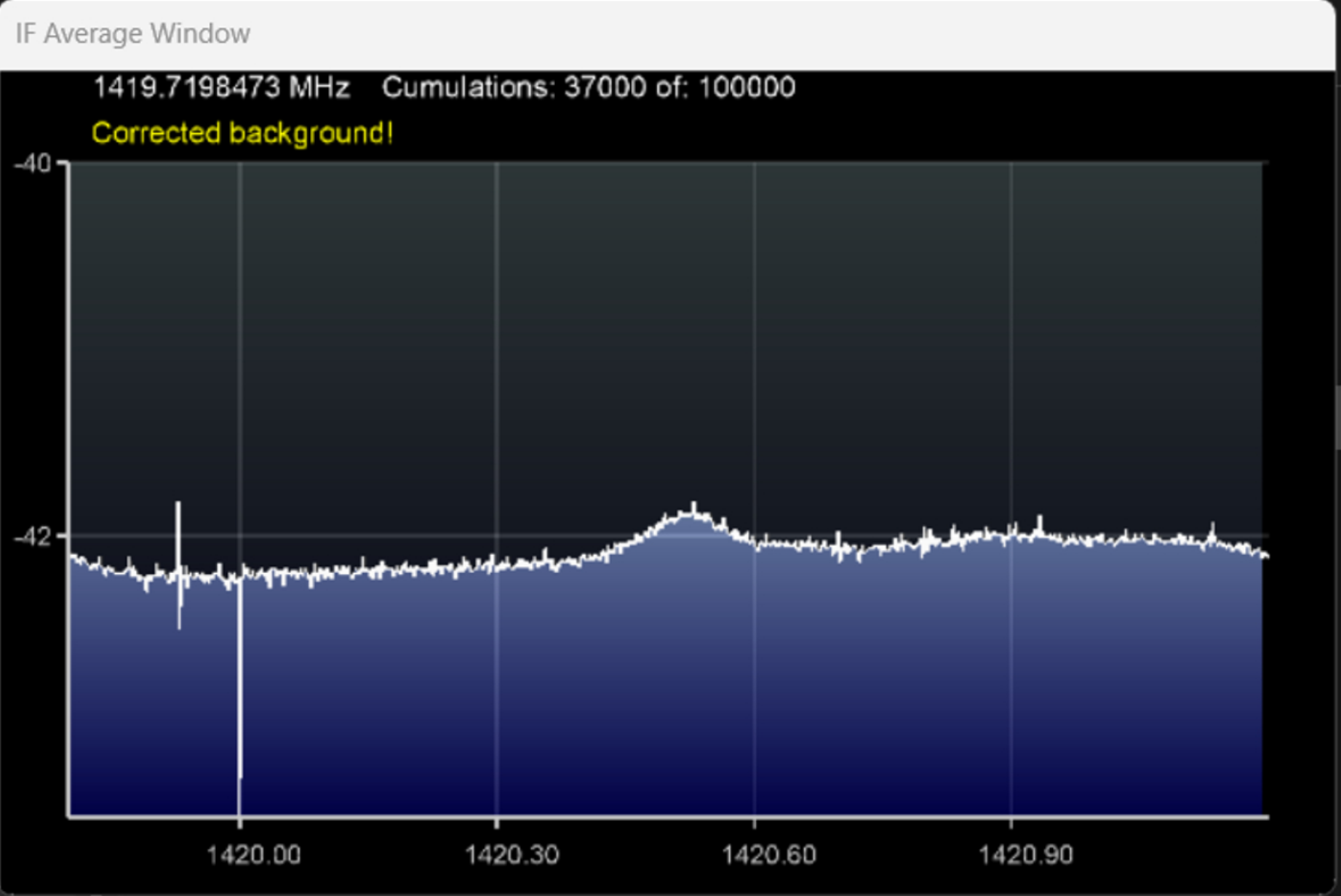}
			\caption{'If Average' plugin window}
			\label{fig:sdr1}
		\end{minipage}
	\end{figure}

		Utilize SDRSharp software to collect data, adjust the parameters of the ‘If Average’ plugin for integral averaging and use background correction to improve signal-to-noise ratio, shown in Figure \ref{fig:sdr1}, and export the spectral data. We can clearly see a prominent peak around 1420.40MHz, which is the hydrogen 21-cm line.\par
	\end{multicols}

\begin{figure}[H]
		\centering
		\begin{minipage}[b]{0.45\textwidth}
			\centering
			\includegraphics[width=\textwidth]{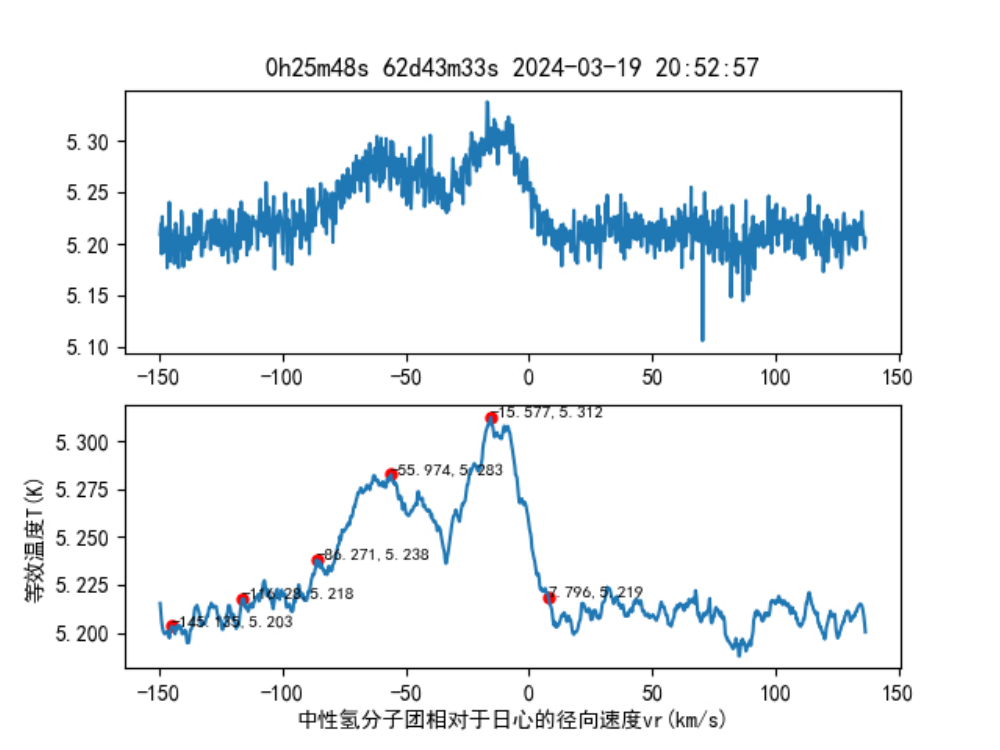}
			\caption{$l=120.0^{\circ}$ spectrum before(uper) and after(lower) smoothing}
			\label{fig:ppsl1}
		\end{minipage}
		\begin{minipage}[b]{0.45\textwidth}
			\centering
			\includegraphics[width=\textwidth]{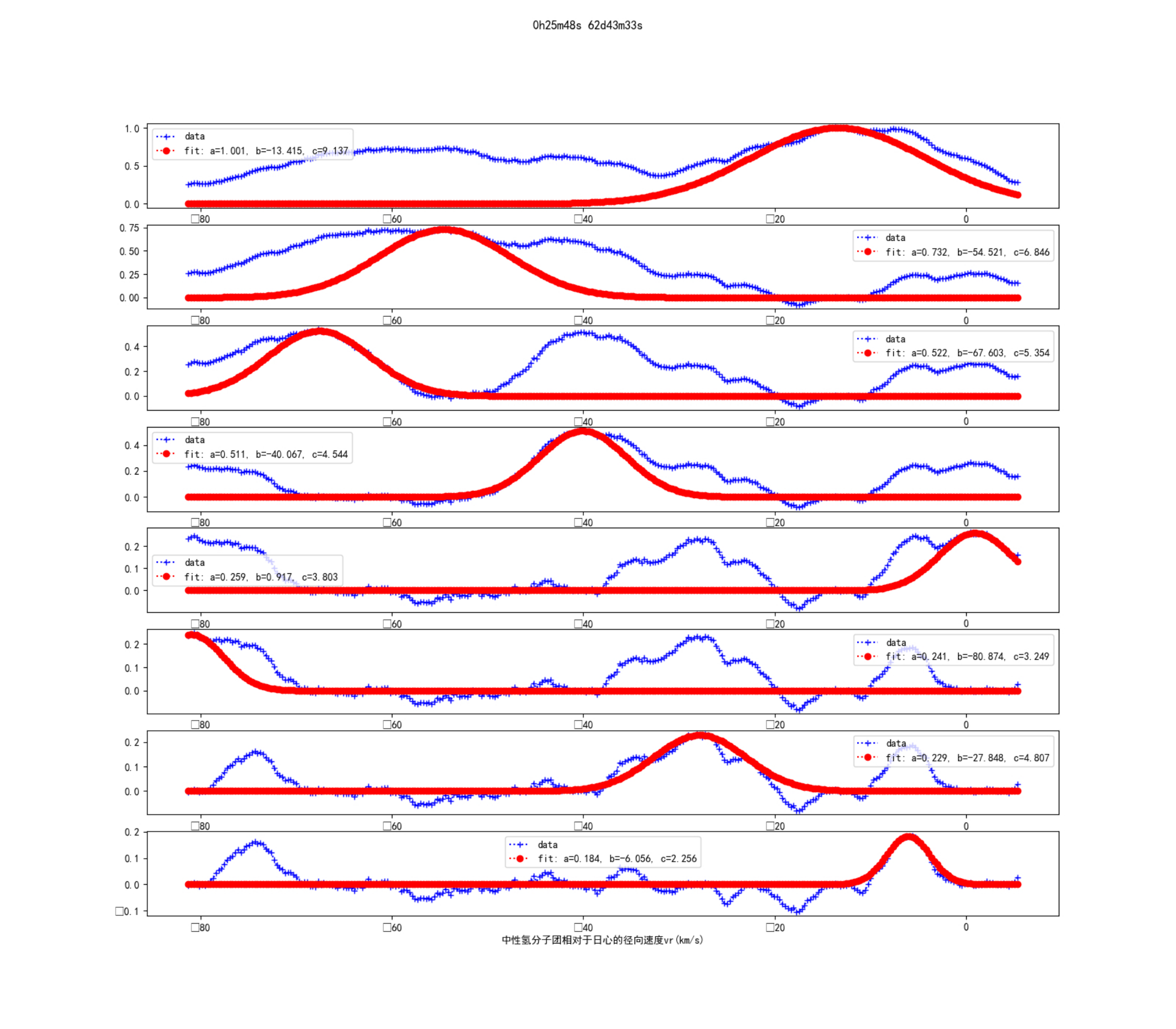}
			\caption{$l=120.0^{\circ}$ spectrum's Gaussian fitting}
			\label{fig:ppsl2}
		\end{minipage}
	\end{figure}

	\begin{multicols}{2}
		We refer the Open Source Project\href{https://github.com/BI6MHT/MilkyWay}{MilkyWay} developed by @BI6MHT to extract our observation data, with the usage of Python. The main steps are as follows:
		\begin{enumerate}
			\item Read the export spectrum data, calculate the corresponding radial velocity and relative equivalent temperature using the Doppler effect, and perform coordinate transformation simultaneously.
			\item Smooth the data to reduce the impact of noise on the data.
			\item Identify peaks (extremum points) to obtain corresponding radial velocities, shown in Figure \ref{fig:ppsl1}.
			\item Fitting Gaussian functions to the peaks to obtain parameters such as position, amplitude, and standard deviation for each peak. To ensure no peak is missed and accuracy, each Gaussian fitting subtracts previous fitted peak data and iterates 8 times, shown in Figure \ref{fig:ppsl2}.
			\item Correct the obtained peaks to radial velocities in the Local Standard of Rest(LSR) reference frame, then convert to positions in the Milky Way galaxy, and plot rotation curves along with Milky Way spiral arm diagrams.
			\item Conduct error analysis.
		\end{enumerate}
		For the spectral data at Galactic longitude(defined as $l$ below) 120°, a clear double-peak structure is evident, with radial velocities less than 0, indicating motion near the Sun. Additionally, besides the hydrogen spectral line data, some noise is observed in the spectrum, which may arise from equipment issues or atmospheric influences. We discard this noise during processing. Regarding Gaussian fitting, slight differences are noted in the fitted parameters compared to those obtained from peak fitting, with the first two peaks being the main ones fitted. Other peaks are considered noise and thus discarded.
	\end{multicols}

	\begin{multicols}{2}
		\section{Rotation Curve and Spiral Structure}
		\subsection{\normalsize{Doppler effect}}
		According to Doppler effect: \[
		\frac{\Delta f}{f_0} = \frac{V_{r}}{c} 
		\]
		where \(\Delta f = f - f_0\) is the frequency shift of observation spectral line, \(f_0\) is stationary frequency, i.e., the 21-cm line's natural frequency, \(V_{r}\) is the radial velocity of neutral hydrogen related to observer, \(c\) is the speed of light. Based on the Doppler effect, the radial velocity of hydrogen molecular clouds in the Earth's reference frame can be calculated, defined as \(V'_{r}\). Additionally, using Python's software package 'astropy' to modify \(V_{r}'\) to \(V_{r}\) in the Sun's reference frame, Frequency spectra can be transformed into radial velocity \(V_{r}\) distribution plots, as shown in Figure \ref{fig:ppsl1}. Simultaneously, the website \href{https://www.gb.nrao.edu/GBT/setups/radvelcalc.html}{Calculate radial velocities of the GBT}\footnote{https://www.gb.nrao.edu/GBT/setups/radvelcalc.html} can convert the radial velocity relative to the Sun to the radial velocity relative to the LSR (defined as \(V_{r}\) below), the radial velocity we need. \par 
		\subsection{\normalsize{Drawing of Rotation Curve}}
\begin{figure}[H]
		\centering
		\includegraphics[width=0.4\linewidth]{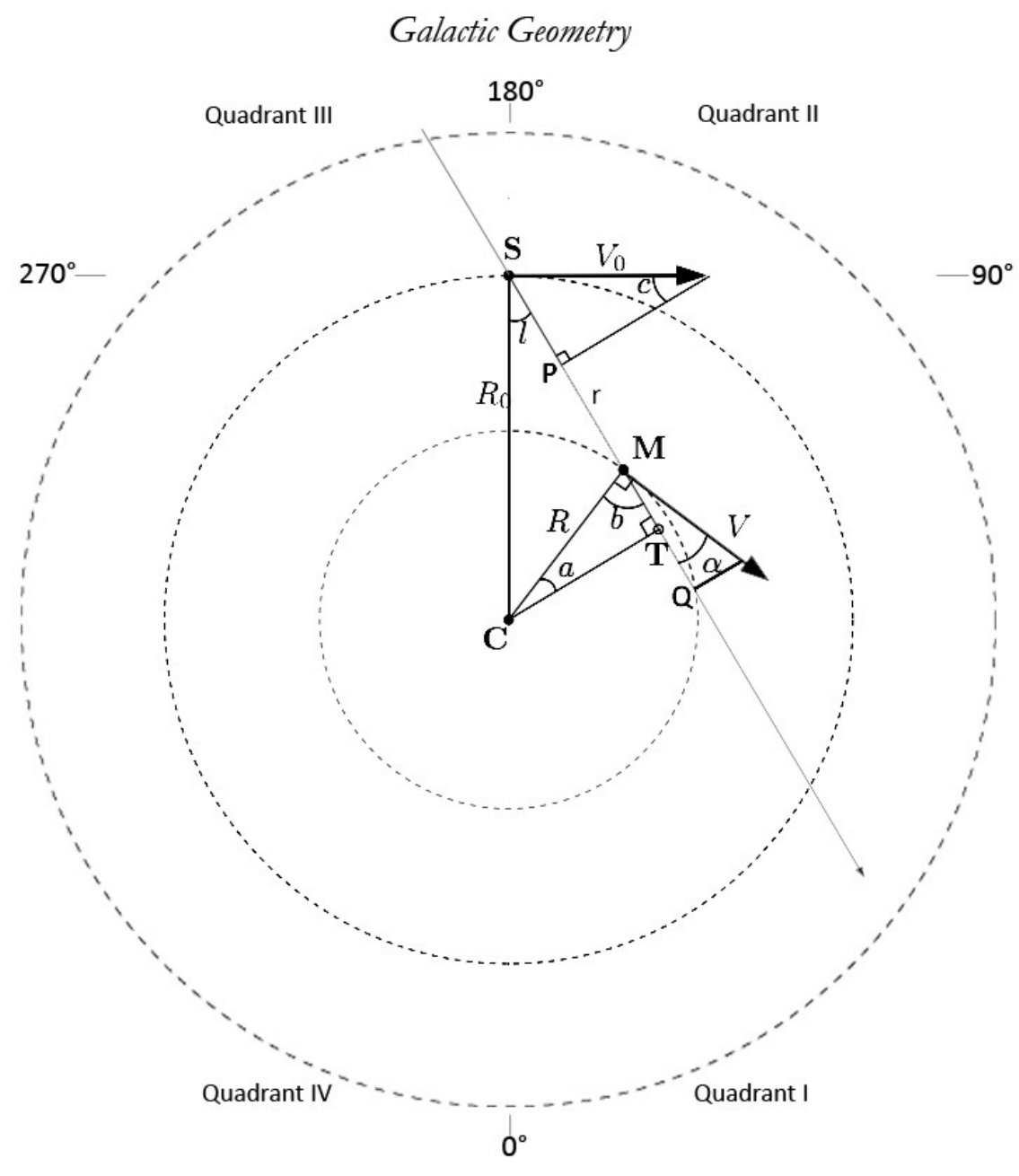}
		\caption{Simplified structure of the Milky Way galaxy}
		\label{fig:structure}
	\end{figure}

		As Figure \ref{fig:structure} show, where \(v(r_0)\) is the rotation speed of the Sun, \(r_0\) is the distance from the Sun to the Galactic center, \(l\) is the observation Galactic longitude, \(r\) is the distance from the observation object to the Galactic center. These parameters satisfy:\Autocite{radioastronomy,salsa}
		\begin{equation}
			V_{r} = v(r)\frac{r_0}{r}\sin l - v(r_0) \sin l \label{eq:vr}
		\end{equation}
		We preliminarily assume that \(v(r) \approx v(r_0) \), and we obtain:
		\begin{equation}
			r \approx \frac{r_0v(r_0) \sin l}{V_{r} + v(r_0) \sin l} \label{eq:R}
		\end{equation}
		From equation \eqref{eq:R} we learn that if the distance between the observation HI cluster and the galactic center takes the minimum value , i.e., \(R_{min }  = r_0 \sin l\), under the fixed observation $l$, the radial velocity \(V_{r}\) will take the maximum value \(V_{rmax}\). Here we take the \(V_{rmax}\) as the radial velocity corresponding to the right foot of the rightmost peak in the spectrum figure, which is indicated by the rightmost red point in the graph. And we can calculate:
		\begin{equation}
			v = V_{rmax } + v(r_0) \sin l
		\end{equation}
		Combining \(v\) with corresponding \(r_{min }\),we draw the rotation curve. The above approximation process essentially proposes the method of tangent line fitting for drawing the rotation curve. However, this approximation may introduce systematic errors, which will be discussed in the next section.\par 
		
		\subsection{\normalsize{Drawing of Spiral Arm}}
		Equipped with rotation curve, i.e., $v(r)$, we rewrite the equation \eqref{eq:vr} into:
		\begin{equation}
			\frac{v(r)}{r}=\frac{1}{r_0}\qty(\frac{V_r}{\sin l}+v(r_0))
		\end{equation}
		The data represented by the red points corresponding to signal peaks indicate the recessional velocities of neutral hydrogen dense regions. By combining the galactic longitude and the rotation curve model, we can calculate the distance from the neutral hydrogen clouds to the galactic center, and thus determine their positions based on geometric relationships. These dense regions of neutral hydrogen correspond to the spiral arm regions of the Milky Way galaxy.Scanning the Galactic plane with a 5° step size in the northern hemisphere’s view, a series of data points of galactic longitude-radius were obtained. Plotting these data points on a Milky Way coordinate chart reveals the local spiral arm structure of the Galaxy prominently.\par 
	\end{multicols}
	
	\begin{multicols}{2}
		\section{\Large{Discussion}}
		\subsection{\normalsize{Comparison of Rotation Curve}}
		Comparing the rotation curve of the galactic disk calculated by us with rotation curve models obtained using different methods in literature, we found that the rotation curve model obtained by us indicates that within the radius of the sun, the rotational speed gradually increases with radius, the rate of increase decreases with radius, and the curve tends towards flat at the radius of the sun. We found that the rotation curve model we obtained does not exhibit a slow decrease in rotational speed with increasing radius near the radius of the sun, as observed by Przemek et al . and Eilers et al.$^{\textcolor{blue}{3}}$, but is instead similar to the model of slow increase tapering off, as proposed by Xin Xiao-Sheng et al.$^{\textcolor{blue}{11}}$ However, at relatively small radii ($<3kpc$), the rotational speed given by our model is significantly lower than that of Xin Xiao-Sheng et al.'s model. When the comparison is narrowed down to models that also use the tangent method, it is found that the models presented by Clemens et al.$^{\textcolor{blue}{4}}$ using CO spectral lines and an article from 2020 that also uses 21cm neutral hydrogen to observe rotation curves$^{\textcolor{blue}{9}}$, both show a slow increase of rotational speed with increasing radius, without a downward trend. Additionally, at radii $<3kpc$, there is also the phenomenon of rotational speed being significantly lower than in Xin Xiao-Sheng et al.'s model.$^{\textcolor{blue}{11}}$ The following discusses and analyzes possible deviations in the observations, and attempts to explain these phenomena.\par

	\end{multicols}
	\begin{figure}[H]
		\centering
		\includegraphics[width=0.9\linewidth]{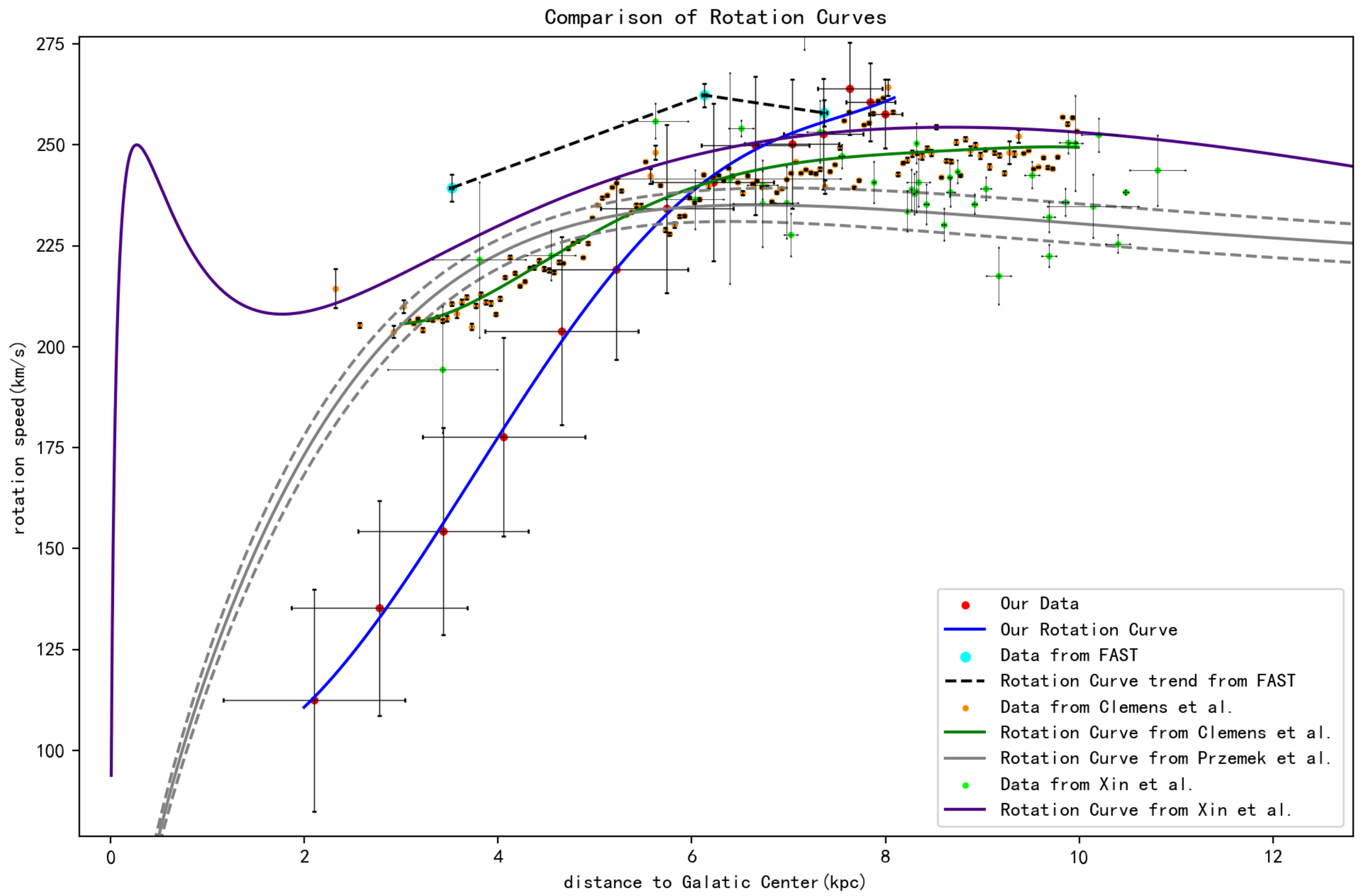}
		\caption{Comparison of Milky Way Galaxy's rotation curve}
		\label{fig:roall3}
	\end{figure}


	
	
	

	
	
	\begin{multicols}{2}
		\subsubsection{Deviations from Peculiar Motions}
		In the rotation curve model, the velocity at a certain radius corresponds to the average speed at which the local material rotates around the galactic center. After subtracting the average speed corresponding to the rotation curve, the velocity residual of the material is described as peculiar motions, i.e., non-circular motions. For the radii of Sun, Przemek et al.$^{\textcolor{blue}{1}}$ pointed out that the rotation speed of  LSR is $233.6\pm2.8$ km/s, and the Sun's peculiar motion, i.e., the motion relate to LSR is $U=11.1\pm1.3$ km/s, $V=12.2\pm2.1$ km/s, $W=7.3\pm0.7$ km/s, where $U$ is the velocity component toward the
		Galactic center, $V$ is the velocity component parallax to the rotation linear velocity of LSR, and $W$ toward the direction of North Galactic pole. For the observed objects at the tangent point, considering the influence of peculiar motion, their line-of-sight recession velocity should be represented by the following equation:

		\begin{equation}
			V_r=v(r)+V-v(r_0)\sin l
		\end{equation}
		We take the maximum redshift of the signal peak of the 21cm neutral hydrogen spectral line as the material with the highest recession velocity relative to us in that direction, and these fastest receding materials beyond the geometric position of the tangent line also have a peculiar motion component moving away from us. Peculiar motions will cause the radial velocity we calculate to be larger than the actual values. Przemek et al.$^{\textcolor{blue}{1}}$ analyzed a large number of Cepheids in the Milky Way and provided a residual map of peculiar motions. We found that the peculiar motions in the Milky Way is generally on the order of 10 km/s, while the rotation curve we observed near the radius of the Sun is about 10-20 km/s larger than the rotational velocity value given by Przemek et al.$^{\textcolor{blue}{1}}$ and others for the LSR.\par 
		\subsubsection{The Problems of Tangent Method}
		When using the tangent method to construct the rotation curve of the Milky Way, we select the maximum red shift of the peak of the hydrogen 21-cm line to represent the maximum recession velocity in that direction. When the line of sight is tangent to the orbit of the observed material orbiting around the Milky Way center, its rotational velocity only has a line-of-sight component, and we believe that the maximum red shift on the spectral line is contributed by these materials. However, when materials on other orbits along the line of sight have rotational velocities significantly greater than those on the tangent orbit, the line-of-sight component of their velocities may be larger than that of the materials on the tangent orbit, and the materials contributing to the maximum red shift may not come from the tangent orbit, causing systematic errors.\par 
		We use geometric relationships to discuss the conditions under which such errors occur and the impact of errors on observational data. When an orbit with radius 
		$r$ is precisely located along the line of sight tangent direction, we have:
		\begin{equation}
			V_{r1}=v(r)-v(r_0)\sin l
		\end{equation}
		For the non-tangent orbit along the same line of sight, the radius of the orbit is taken as $r+\Delta r$, the angle between the rotation velocity and the line of sight is $\theta$, and we have:
		\begin{equation}
			V_{r2}=v(r+\Delta r)\cos \theta-v(r_0)\sin l
		\end{equation}
		If $V_{r2}>V_{r1}$, there is a systematic error in tangent method. Order the value $\Delta r\rightarrow0$ and refer to derivative definition, we learn the prerequisite for the systematic errors from tangent method corresponding to radius $r$ is $v^{'}(r)>\frac{v(r)}{r}$. In other words, when the slope at a point on the rotation curve is greater than the slope of the line tangent to that point and the origin, the material contributing to the maximum red shift at that point does not come from the tangent orbit. In reality, the maximum red shift comes from radius $\overline{r}$, and  rotation speed $\overline{v}$. They show a relation with the corresponding parameters of tangent point:
		\begin{equation}
			\overline{r}=r+\Delta r
		\end{equation}
		\begin{equation}
			\begin{array}{cl}
				\overline{v}=v(r+\Delta r)\approx v(r)+v^{'}(r)\Delta r  \\> v(r)+\frac{v(r)}{r}\Delta r=\overline{\overline{v}}
			\end{array}
		\end{equation}
		The extension of the line connecting the origin and $(r,v)$ precisely goes through $(\overline{r},\overline{\overline{v}})$. It can be inferred that the slope of the line connecting the origin and $(\overline{r},\overline{v})$ is greater than that of the former. Analyzing the rotation curve, it can be observed that the slope of the rotation curve is larger at smaller radii positions, making it susceptible to this systematic error, resulting in a lower measured rotation curve velocity. This can be used to qualitatively explain the phenomenon where the rotation speed in the tangent method is lower than the model proposed by Xin Xiao-Sheng et al. for radii <3kpc.\par 
		\subsubsection{Comparing with Data from FAST}
		For further comparison, we utilized the neutral hydrogen survey data from FAST \href{http://groups.bao.ac.cn/ism/CRAFTS/202309/t20230906_752058.html}{CRAFTS Narrow-band data-cube DR1}\footnote{http://groups.bao.ac.cn/ism/CRAFTS/202309/t20230906\_752058.html} and extracted three sets of neutral hydrogen spectral data located in the direction of the Galactic disk and similarly obtained the corresponding radius and rotational velocity using the tangent method. The large aperture of FAST provides high resolution and signal-to-noise ratio, hence the accuracy of its data is unquestionable. From Figure \ref{fig:roall3}, it can be observed that the velocity values are significantly higher, indicating the widespread presence of peculiar motions. However, at smaller radii, the data from FAST does not show as significant a decrease as our observational data, which could be due to our signal-to-noise ratio not being high enough, causing signal peaks to be submerged in noise, resulting in a smaller maximum recession velocity being read. In terms of the shape of the rotation curve, the observational data from FAST gradually declines near the radius where the Sun is located, which is different from the conclusion drawn using the tangent method by us and Clemens et al.$^{\textcolor{blue}{4}}$ The shape of the rotation curve from FAST is more similar to the model proposed by Przemek et al.$^{\textcolor{blue}{1}}$\par 
		\subsection{\normalsize{Map Material Distribution}}
		In Section 3, we present a method for determining the positions of high-density neutral hydrogen clouds in the Milky Way using red shift values obtained from the peak of the hydrogen 21-cm line. Since hydrogen constitutes the majority of elements in the universe, we assume that the density distribution of neutral hydrogen clouds is positively correlated with the matter density distribution, yielding a map of the spiral arms of the Milky Way. Due to limitations in observing from the southern hemisphere and the  
		$1/\sin l$ effect, we only depict a local map of the spiral arm distribution, which clearly shows the local structure of the three spiral arms of the Milky Way. We compare spiral arm distribution maps based on different rotation curve models, discuss phenomena encountered during data processing, and deepen our understanding of rotation curves and spiral arm distribution.\par 
		
	\end{multicols}
	\begin{figure}[H]

		\begin{minipage}[b]{0.49\textwidth}        
			\centering          
			\includegraphics[width=0.95\textwidth]{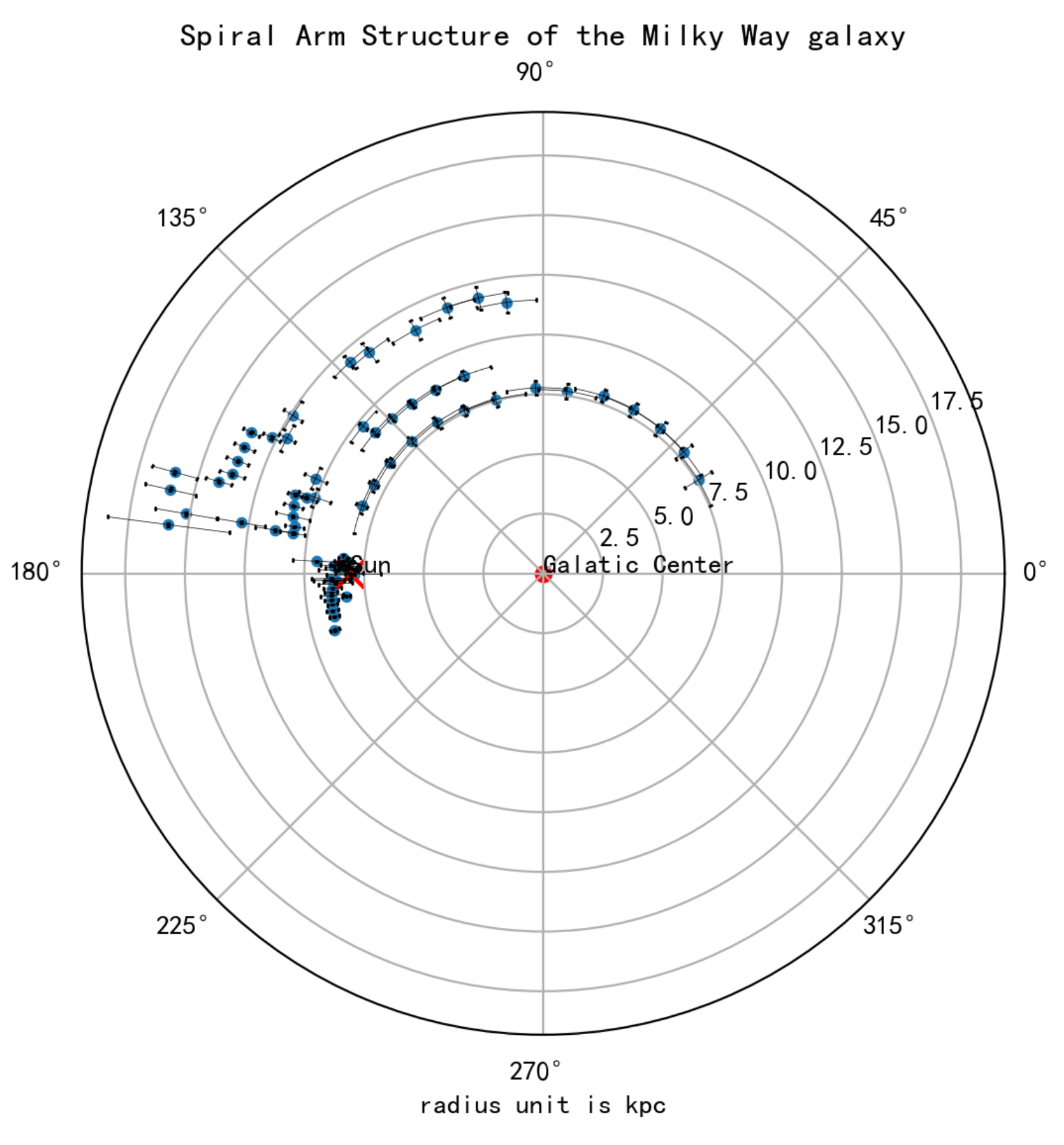}
			\caption[ ]{Based on Przemek et al.'s model}
			
		\end{minipage}
		\begin{minipage}[b]{0.49\textwidth}
			\centering
			\includegraphics[width=0.95\textwidth]{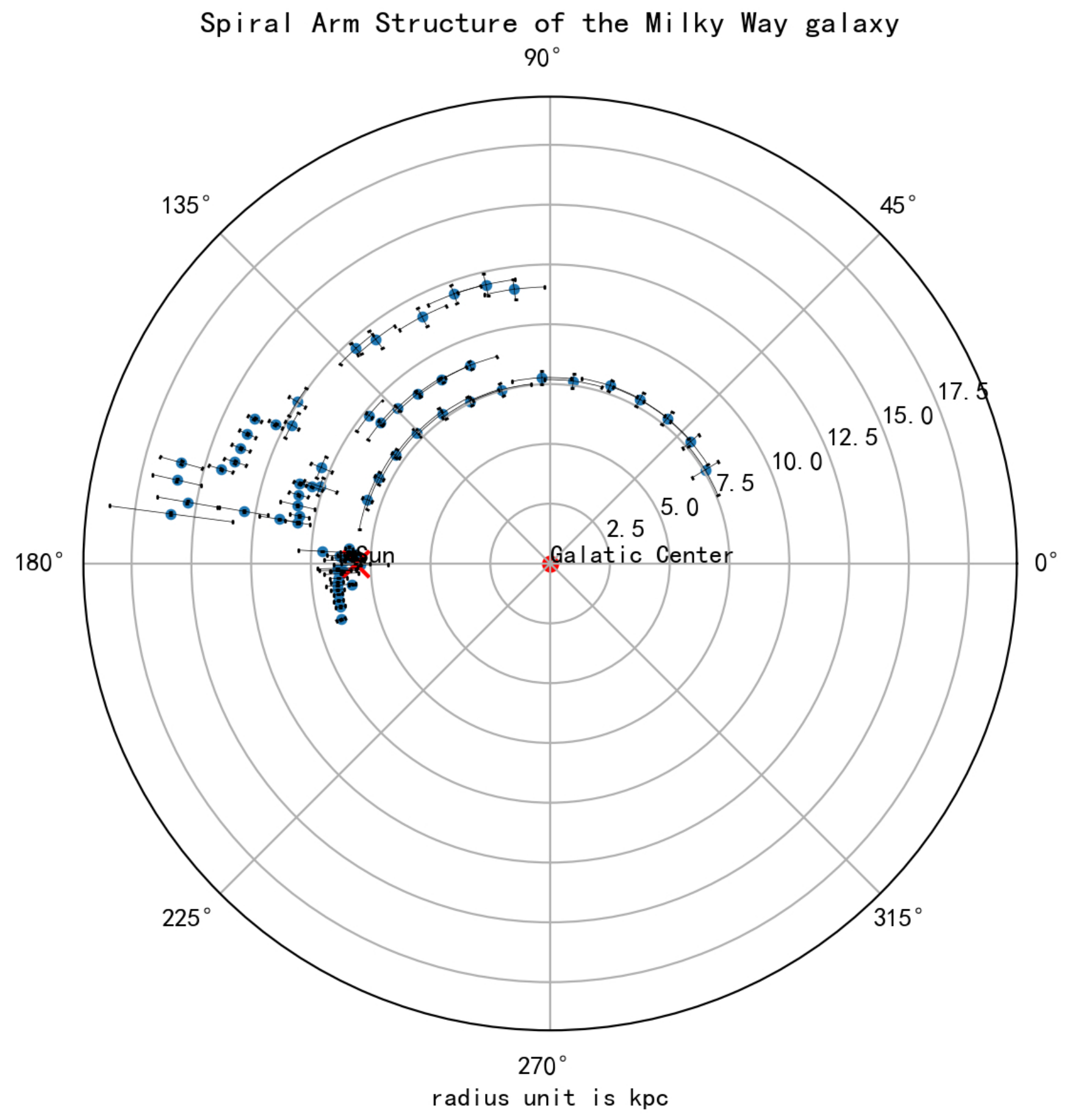}
			\caption[ ]{Based on Eilers et al.'s model}
			
		\end{minipage}
	\end{figure}
	\begin{figure}[H]

		\begin{minipage}[b]{0.49\textwidth}        
			\centering          
			\includegraphics[width=0.95\textwidth]{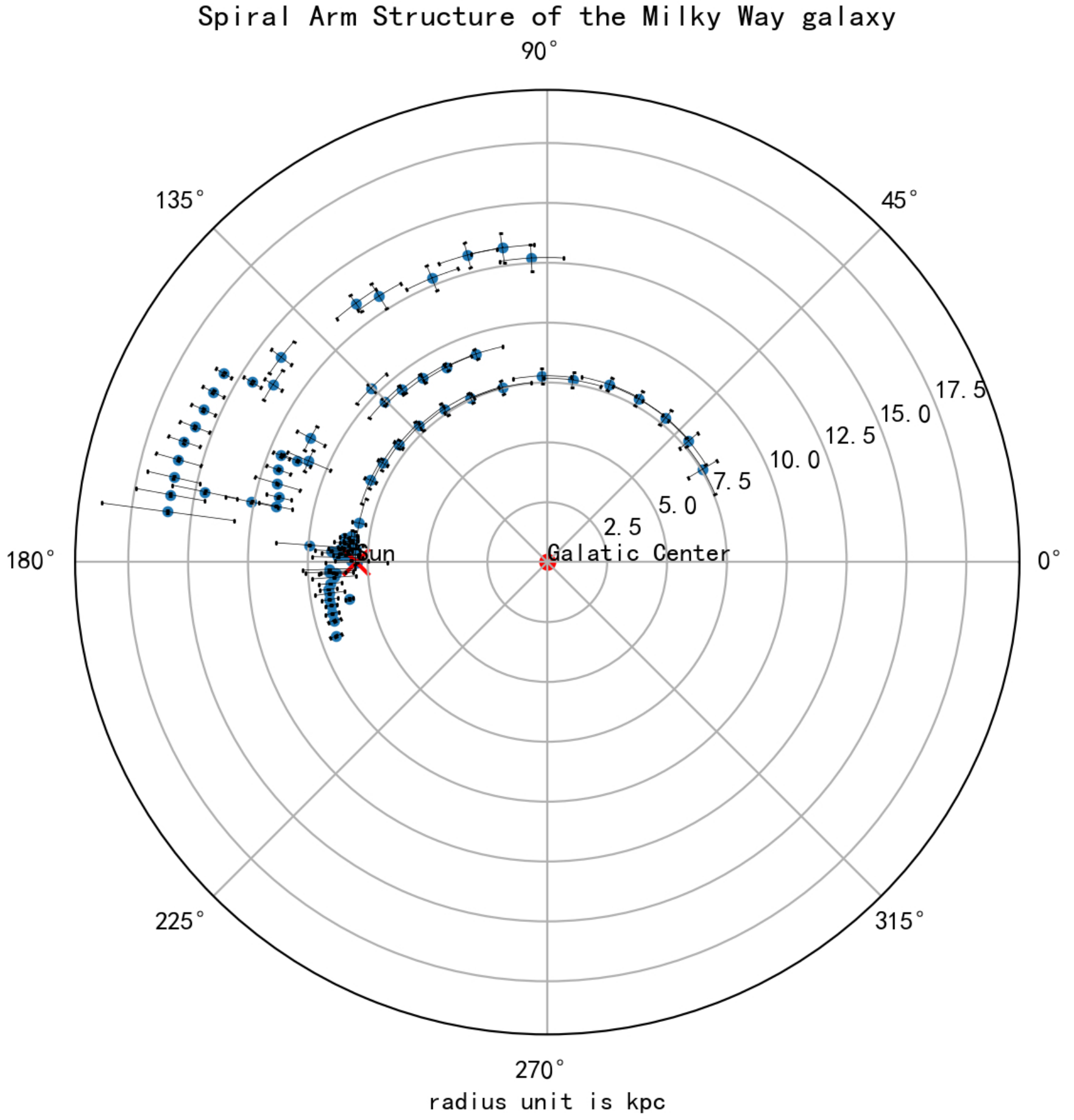}
			\caption[ ]{Based on Sofue et al.'s model}
			
		\end{minipage}
		\begin{minipage}[b]{0.49\textwidth}
			\centering
			\includegraphics[width=0.95\textwidth]{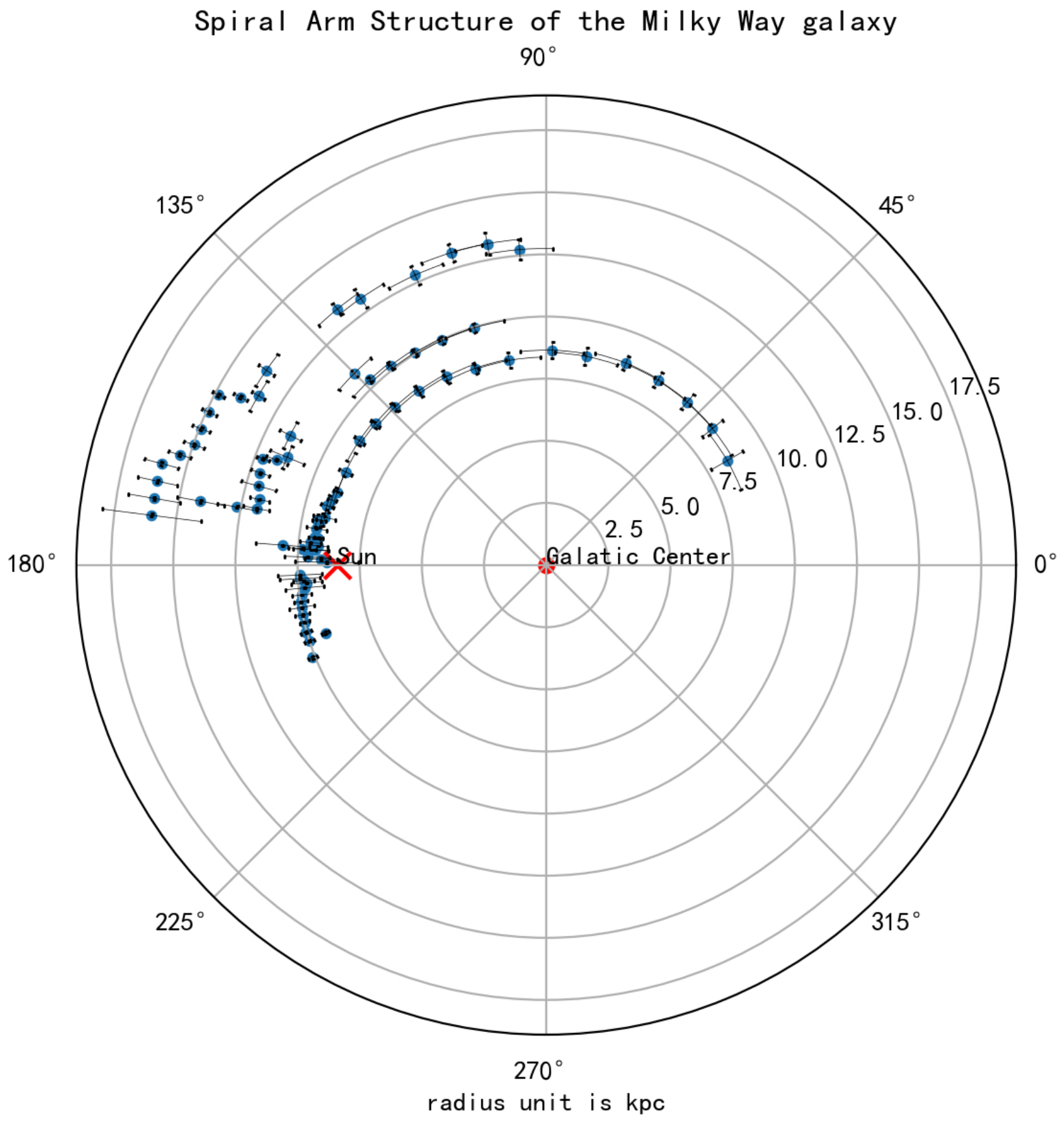}
			\caption[ ]{Based on Xin et al.'s model}
			
		\end{minipage}
	\end{figure}

	\begin{multicols}{2}
		\subsubsection{$1/\sin{l}$ effect}
		By peak corresponding red shift values, combined with the existing rotation curve model $v(r)$, we can determine, through geometric relationships, the distance from the signal peak position to the Galactic center corresponding to regions of higher-density hydrogen clouds, as expressed in the preceding text:
		\begin{equation}
			\frac{v(r)}{r}=\frac{1}{r_0}\qty(\frac{V_r}{\sin l}+v(r_0))
		\end{equation}
		The Galactic longitude $l$ is observed quantity. Considering the right-hand side of the equation, the error component it introduces is $|\frac{V_r\cos l}{r_0\sin^2l}|\Delta l$. From this, it can be seen that when 
		$\sin l$ is relatively small, the error propagation coefficient becomes significant, leading to large uncertainties in the distance from neutral hydrogen clouds to the Galactic center. As a result, we disregard the data points that fall within $0^{\circ} < l< 15^{\circ},\quad 165^{\circ}< l <195^{\circ},\quad 345^{\circ}< l<360^{\circ}$. This operation results in two ‘blind zones’ of 30° each towards and away from the Galactic center. From the diagram of our Milky Way’s spiral arm structure, it can be observed that data points closer to these two directions exhibit longer radial error bars, demonstrating the impact of the 
		$1/\sin l$
		effect on our observational calculations. This phenomenon has also been mentioned in Sofue et al.'s article.\textcolor{blue}{$^{10}$}\par

		\subsubsection{The Deviation near $l=90^{\circ}$}
		When representing the positions of high-density regions of neutral hydrogen clouds in polar coordinates with the galactic center as the origin, it is necessary to obtain the distance $\tilde{r}$
		of neutral hydrogen clouds to the Sun, and then calculate their coordinates through geometric relationships. It satisfies the equation:
		\begin{equation}
			\tilde{r}=r_0\cos l\pm\sqrt{r_0^2\cos^2 l-(r_0^2-r^2)}
		\end{equation}
		For the outer disk, only the positive sign can satisfy the physical reality that the distance is always greater than 0. For the inner disk, there are two solutions for its position. However, when we process the data, we find that some individual data points $(l,r)$ result in situations where there is no solution, i.e., less than zero under the root sign. We found that this type of unsolvable data points occurs around $l=90^{\circ}$ when applying the rotation curve models by Przemek et al.$^{\textcolor{blue}{1}}$ and Eilers et al.$^{\textcolor{blue}{3}}$ From the formula, it is easy to see that the numerical value of the first term under the square root is relatively small at these data points. As long as the second term generates a small negative value, it will lead to no solution. Furthermore, considering physical reality, the condition for solvability around $l=90^{\circ}$ is:
		\begin{equation}
			\left\{
			\begin{array}{cl}
				r> r_0\sin l,\quad l< 90^{\circ}\\
				r> r_0,\quad l\geq 90^{\circ}\\
			\end{array}
			\right.		
		\end{equation}
		According to the obtained spiral arm diagram, we believe that our observed objects revolve around a radius that differs from 
		$r_0$
		by a small amount, i.e., 
		$r-r_0\rightarrow 0$. Using a first-order Taylor expansion, we obtain:
		\begin{equation}
			(\frac{v(r)}{r})^{'}|_{r=r_0}\approx\frac{\frac{v(r)}{r}-\frac{v(r_0)}{r_0}}{r-r_0}=\frac{V_r}{r_0(r-r_0)\sin l}
		\end{equation}
		Further:
		\begin{equation}
			\left\{
			\begin{array}{cl}
				V_r<(\frac{v(r)}{r})^{'}|_{r=r_0}r_0^2\sin l(\sin l-1) ,\quad l< 90^{\circ}\\
				V_r<0,\quad l\geq 90^{\circ}\\
			\end{array}
			\right.		
		\end{equation}
		Considered data from several rotation curve models, we evaluate $(\frac{v(r)}{r})^{'}|_{r=r_0}\approx-3km*s^{-1}*kpc^{-2}$ and $r_0\approx8kpc$, We find that when $l$ go closer to $90^{\circ}$, the upper limit of $V_r$ decreases rapidly. We take some special values for $l$ and see that:
		\begin{equation}
			\left\{
			\begin{array}{cl}
				V_r <6.97km*s^{-1}\quad,\quad l=75^{\circ}\\
				V_r <3.17km*s^{-1}\quad,\quad l=80^{\circ}\\
			\end{array}
			\right.		
		\end{equation}
		If the observed object is located in the inner disk, it will satisfy 
		$V_r>0$, when $l$ approaches $90^{\circ}$, its recession velocity satisfies the narrow “bandwidth” corresponding to the solvability condition. However, the $1\sigma$ error of the observed recession velocity $V_r>0$ is around $5km*s^{-1}$, so there is a considerable probability that the recession velocity observed when $l>80^{\circ}$
		falls outside the solvable range due to observational errors.Similarly, for the outer disk portion near $90^{\circ}$, it is also possible that observational errors may lead to observed $V_r>0$ causing unsolvability. Due to no solution situation that prevent further computation and plotting, we had to remove data points resulting in unsolvable outcomes according to the rotation curve models by Przemek et al.$^{\textcolor{blue}{1}}$ and Eilers et al.$^{\textcolor{blue}{3}}$ For the models by Sofue et al.$^{\textcolor{blue}{10}}$ and Xin Xiao-Sheng et al.$^{\textcolor{blue}{11}}$, no unsolvable outcomes were generated, and all data could be used for plotting. Combining the spiral arm distribution we constructed, using the models by Przemek et al.$^{\textcolor{blue}{1}}$ and Eilers et al.$^{\textcolor{blue}{3}}$, we conclude that the Sun is located on the Local Arm; however, observational data around $l=90^{\circ}$ on the Local Arm yielded unsolvable results due to measurement errors. In contrast, using the models by Sofue et al.$^{\textcolor{blue}{10}}$ and Xin Xiao-Sheng et al.$^{\textcolor{blue}{11}}$, we propose that the Sun resides on the inner side of the Local Arm, with the distance from $l=90^{\circ}$ on the Local Arm to the Galactic center significantly greater than that of the Sun, resulting in smaller recession velocities and hence solvable data. The unsolvability near $l=90^{\circ}$ can be attributed to deviations caused by observational errors, thus supporting all the rotation curve models mentioned above with our observations, albeit yielding slightly different distributions of spiral arms.\par 
		\subsection{\normalsize{Compare with Data from LAB}}
		To verify the accuracy of the spectrum obtained from our observations using a small radio telescope, we cross-referenced the data with the LAB neutral hydrogen survey:\href{https://lambda.gsfc.nasa.gov/product/foreground/fg_LAB_HI_Survey_info.html}{LAB\_HI\_Survey Maps}\footnote{https://lambda.gsfc.nasa.gov/product/foreground/fg\_LAB\_HI\_Survey\_info.html}. And we extracted neutral hydrogen spectral data in different directions on the Galactic plane using the angular resolution corresponding to the aperture of our small radio telescope, comparing and analyzing it against our obtained spectral data. In the Local Standard of Rest (LSR) reference frame, the velocities corresponding to the peaks in both datasets are nearly identical, indicating the accuracy and reliability of the peak position data observed by our telescope.\par 
	\end{multicols}
	\begin{figure}[H]

		\begin{minipage}[b]{0.49\textwidth}        
			\centering          
			\includegraphics[width=0.95\textwidth]{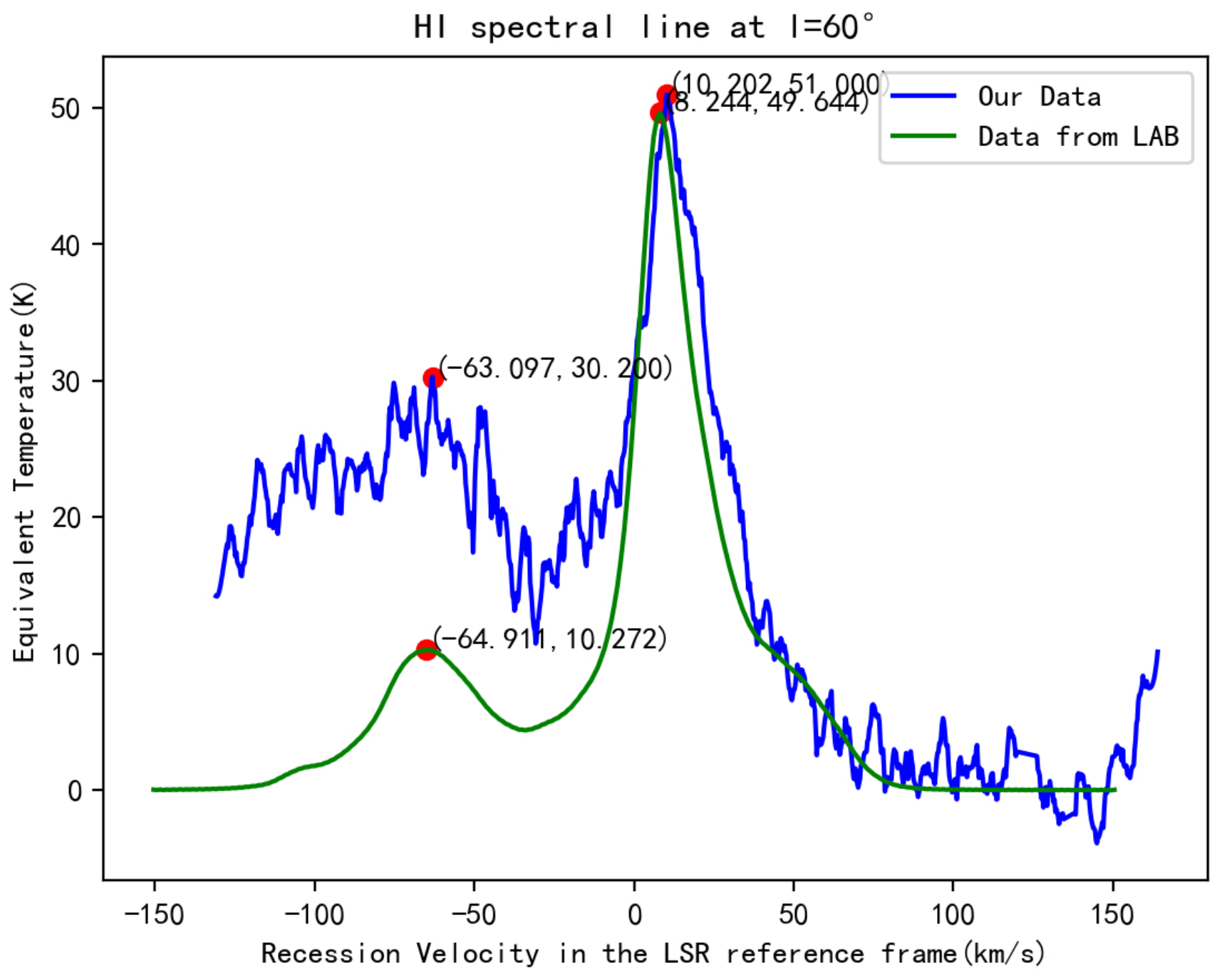}
			\caption[ ]{Data Comparison at $l=60^{\circ}$}
			\label{fig:60}
		\end{minipage}
		\begin{minipage}[b]{0.49\textwidth}
			\centering
			\includegraphics[width=0.95\textwidth]{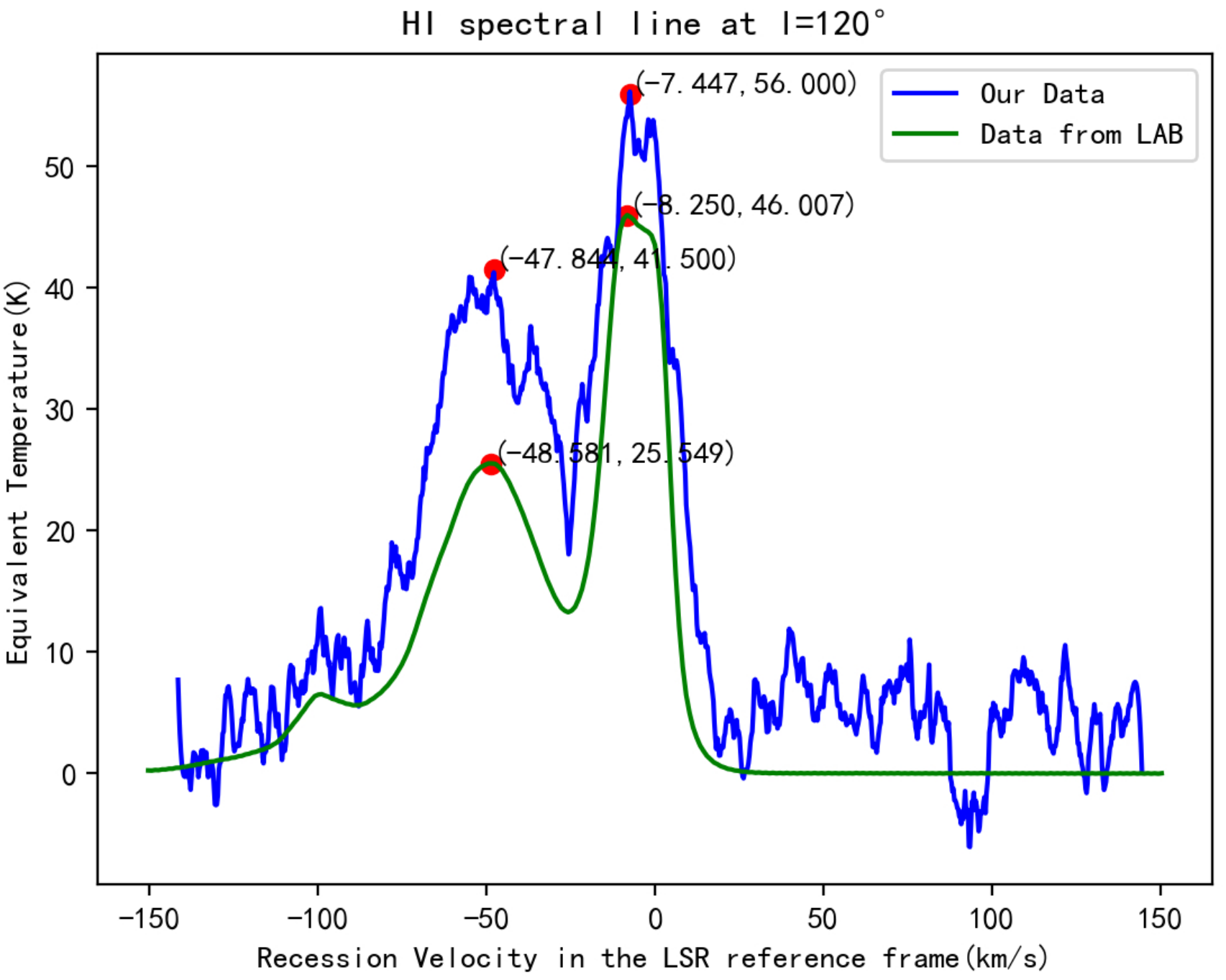}
			\caption[ ]{Data Comparison at $l=120^{\circ}$}
			
		\end{minipage}
	\end{figure}

	\begin{multicols}{2}
		\section{Conclusion}
		We conducted radio astronomy observations on campus using a small radio antenna and receiver, aiming to observe the 21cm neutral hydrogen emission line on the Galactic plane. Utilizing the position and velocity of the Sun in the Galactic frame provided by Przemek et al.$^{\textcolor{blue}{1}}$, we employed the tangent method to plot the rotation curve of the inner Milky Way disk. The rotation curve exhibits a steadily rising trend that tends to flatten out. Furthermore, based on existing rotation curve models in the astronomical community, we further mapped out the local spiral arm distribution of the Milky Way, revealing three distinct arm structures. Through analysis of observations and comparisons, we discussed potential systematic errors in the tangent method for plotting rotation curves, explained the physical interpretations corresponding to unsolvable situations encountered during the mapping of spiral arm distribution, and further identified conclusions regarding the Sun’s position on the Local Arm using the models by Przemek et al.$^{\textcolor{blue}{1}}$ and Eilers et al.$^{\textcolor{blue}{3}}$, and on the inner side of the Local Arm using the models by Sofue et al.$^{\textcolor{blue}{10}}$ and Xin Xiao-Sheng et al.$^{\textcolor{blue}{11}}$ We attribute unsolvable situations during data processing to observational errors and assert that the rotation curve models used for plotting the spiral arm distribution are supported by our observations.\par
		We only corrected for the motion of the Sun relative to the Local Standard of Rest, making the restrictive assumption that material at a certain radius orbits the Galactic center at the same speed, thus neglecting peculiar motions, resulting in a rotation curve higher than actual. Additionally, the steeper slope in the rotation curve at smaller radii  may causes the maximum red shift not to be solely contributed by the motion of material at the tangent point, leading to a lower-than-actual rotation curve at smaller radii. During the process of constructing the spiral arm distribution map, influenced by the $1/sinl$ effect, we had to discard observational data near Galactic longitudes 0° and 180°, and due to observational errors of the small radio telescope, deviations in data around Galactic longitude 90° resulted in unsolvable situations during data processing.\par 
		Our work demonstrates the feasibility of conducting astronomical observations on university campuses, providing valuable insights for undergraduate astronomical projects and the cultivation of talent in university astronomy programs. However, there is ample room for improvement in observation equipment, observation strategies, model assumptions, and data processing. In the future, building upon the preliminary results of this project, we aim to construct more accurate rotation curves and spiral arm distribution maps.\par 
		\section{Acknowledgments}
		Special thanks to the support from the university’s innovation and entrepreneurship program. Since the initiation of the project, we extend our gratitude to Professor Weihua Lei from the Department of Astronomy for his valuable guidance and suggestions. We also appreciate the active participation of our team members, students  Jiahao Hu and Qishuo Zhang, throughout the project. Additionally, we would like to thank Professor Tang Ningyu from Anhui Normal University for providing the FAST and LAB survey data, as well as for offering guidance and advice on data processing. Our thanks also go to postgraduate student Po Ma for the patient guidance and assistance with installing and using the astronomical data processing software DS9.
		\nocite{*}
		\printbibliography

@article{liu2008hydrogen,
  title={The hydrogen 21-cm line and its applications to radio astrophysics},
  author={Liu, Lulu},
  journal={Massachusetts Institute of Technology},
  year={2008}
}

@online{radioastronomy,
  author    = {},
  title     = {Amateur Radio Astronomy Handbook},
  year      = {},
  url       = {https://bi6mht.gitee.io/radioastronomy/#/},
  urldate     = {2024-04-10}
}

@online{salsa,
  title     = {SALSA-HI English},
  url       = {https://raw.githubusercontent.com/varenius/salsa/main/Lab_instructions/HI/English/SALSA-HI_English.pdf},
  urldate   = {2024-04-10}
}

@article{mroz2019rotation,
  title={Rotation curve of the milky way from classical cepheids},
  author={Mr{\'o}z, Przemek and Udalski, Andrzej and Skowron, Dorota M and Skowron, Jan and Soszy{\'n}ski, Igor and Pietrukowicz, Pawe{\l} and Szyma{\'n}ski, Micha{\l} K and Poleski, Rados{\l}aw and Koz{\l}owski, Szymon and Ulaczyk, Krzysztof},
  journal={The Astrophysical Journal Letters},
  volume={870},
  number={1},
  pages={L10},
  year={2019},
  publisher={IOP Publishing}
}

@article{clemens1985massachusetts,
  title={Massachusetts-Stony Brook Galactic plane CO survey-The Galactic disk rotation curve},
  author={Clemens, Dan P},
  journal={Astrophysical Journal, Part 1 (ISSN 0004-637X), vol. 295, Aug. 15, 1985, p. 422-428, 431-436.},
  volume={295},
  pages={422--428},
  year={1985}
}

@article{mhaske2022bose,
  title={A Bose horn antenna radio telescope (BHARAT) design for 21 cm hydrogen line experiments for radio astronomy teaching},
  author={Mhaske, Ashish A and Bagchi, Joydeep and Joshi, Bhal Chandra and Jacob, Joe and KT, Paul},
  journal={American Journal of Physics},
  volume={90},
  number={12},
  pages={948--960},
  year={2022},
  publisher={AIP Publishing}
}

@article{eilers2019circular,
  title={The circular velocity curve of the milky way from 5 to 25 kpc},
  author={Eilers, Anna-Christina and Hogg, David W and Rix, Hans-Walter and Ness, Melissa K},
  journal={The Astrophysical Journal},
  volume={871},
  number={1},
  pages={120},
  year={2019},
  publisher={IOP Publishing}
}

@online{milkywayrotation,
  title     = {Measurement of the Milky Way Rotation},
  url       = {https://physicsopenlab.org/2020/09/08/measurement-of-the-milky-way-rotation/},
  urldate   = {2024-04-10}
}

@article{sofue2009unified,
  title={Unified Rotation Curve of the Galaxy—Decomposition into de Vaucouleurs Bulge, Disk, Dark Halo, and the 9-kpc Rotation Dip—},
  author={Sofue, Yoshiaki and Honma, Mareki and Omodaka, Toshihiro},
  journal={Publications of the Astronomical Society of Japan},
  volume={61},
  number={2},
  pages={227--236},
  year={2009},
  publisher={Oxford University Press}
}

@article{reid2014trigonometric,
  title={Trigonometric parallaxes of high mass star forming regions: the structure and kinematics of the Milky Way},
  author={Reid, MJ and Menten, KM and Brunthaler, A and Zheng, XW and Dame, TM and Xu, Y and Wu, Y and Zhang, B and Sanna, A and Sato, M and others},
  journal={The Astrophysical Journal},
  volume={783},
  number={2},
  pages={130},
  year={2014},
  publisher={IOP Publishing}
}

@article{xin2013revised,
  title={A revised rotation curve of the Milky Way with maser astrometry},
  author={Xin, Xiao-Sheng and Zheng, Xing-Wu},
  journal={Research in Astronomy and Astrophysics},
  volume={13},
  number={7},
  pages={849},
  year={2013},
  publisher={IOP Publishing}
}
		
	\end{multicols}
\end{document}